# Child Education Through Animation: An Experimental Study

Md. Baharul Islam[1], Arif Ahmed[2], Md. Kabirul Islam[2] and Abu Kalam Shamsuddin[2]

[1]Faculty of Computing and Informatics, Multimedia University, Cyberjaya, Malaysia
[2]Department of Multimedia and Creative Technology, Daffodil International University, Dhaka-1207, Bangladesh

## *ABSTRACT*

*Teachers have tried to teach their students by introducing text books along with verbal instructions in traditional education system. However, teaching and learning methods could be changed for developing Information and Communication Technology (ICT). It's time to adapt students with interactive learning system so that they can improve their learning, catching, and memorizing capabilities. It is indispensable to create high quality and realistic leaning environment for students. Visual learning can be easier to understand and deal with their learning. We developed visual learning materials (an overview of solar system) in the form of video for students of primary level using different multimedia application tools. The objective of this paper is to examine the impact of student's abilities to acquire new knowledge or skills through visual learning materials and blended leaning that is integration of visual learning materials with teacher's instructions. We visited a primary school in Dhaka city for this study and conducted teaching with three different groups of students (i) teacher taught students by traditional system on same materials and marked level of student's ability to adapt by a set of questions (ii) another group was taught with only visual learning material and assessment was done with 15 questionnaires, (iii) the third group was taught with the video of solar system combined with teacher's instructions and assessed with the same questionnaires. This integration of visual materials (solar system) with verbal instructions is a blended approach of learning. The interactive blended approach greatly promoted students ability of acquisition of knowledge and skills. Students response and perception were very positive towards the blended technique than the other two methods. This interactive blending leaning system may be an appropriate method especially for school children.*

## 1. INTRODUCTION

Interactive learning means compute-based learning system which response to the students' actions by presenting contents such as texts, graphics, animation, video, audio etc [1]. We have chance to change our traditional reading and memorization habits with interesting contents by effective use of technology. Different communication media have promoted implementation of this technology. Learning environment is the big factor to adapt students with their learning system. Multimedia Technology can help to create high quality learning environments especially for students through , different medias like texts, graphics, sound, animation etc. It is true that traditional education is slowly moving away from pen-and-paper correspondence courses, allowing for a more interactive, integrated learning environment [2]. The term *'blended learning'* has gained considerable attention in recent years as particular forms of teaching with technology. The combination of traditional learning with media and tools are employed as blended learning. The Department for Education and Skills in 2003 [3], for example, provided the following definition:

"If someone is learning in a way that uses information and communication technologies, they are using e-learning. They could be a pre-school child playing an interactive game; they could be a group of pupils collaborating on a history project with pupils in another country via the Internet – it all counts as e-learning."

Actually due to the disadvantages of E-Learning a new approach 'blended learning' has been developed [4]. The purpose of blended learning is to discover the mixing of media [5] in education system. The main focus of this research will be to determine most effective learning methods. Our study will observe to test the following hypotheses:

a. The effectiveness of visual learning when integrated with a traditional learning

b. To observe students performance after learning blended education mode

The basic concept of classroom based tuition will be combined with interactive multimedia resources. Blended learning comes in many shapes and types. It may be used to enhance the traditional lecture with additional readings, electronic instructor notes and images of charts, graphs, or other handouts in one course. Indeed, teaching is the passion and relationship between the teacher and the student. Technology is seen as being possibly useful in supporting face-to-face teaching, enabling students to interact with learning material [4]. Main purpose of this study is to investigate primary level students' responses on multimedia contents in learning geography (solar system) with animated models and information. This paper is organized following by literature review, methodology, result analysis, and conclusion.

## 2. LITERATURE REVIEW

Modern education and communication environments can offer alternative ways in the learning process. Technology has been widely used in educational technologies. It is very near when multimedia tools will be perfect utilization in education sector. Using interactive multimedia in the teaching process is growing in the present context. Interactive learning plays a very important role in assisting students in learning processes [6]. It has possibility to enhance the early education system with multimedia technologies. The positive impact of the developed program on students abilities is to understand new knowledge or skills. Multimedia education offers an alternative to traditional education that can enhance the current methods and provide an alternative. Similarly, another group of researchers [7] developed an English short play as a teaching material to promote children's (second language learners) English learning attitude and interests that was presented to all classmates and evaluated by three professors. The findings of the study reveal that incorporating project-based learning into the development of an English short play can effectively guide students in creating the short play effectively. Animation media can help children to adopt their English vocabulary and receive higher average score than those who apply the normal one at statistical significance level [8]. It is a beneficial teaching material to stimulate and support the learners, especially at 5 to 6 years old to enjoy the class with good results.

Some work conducted at Sri Lankan universities in integrating web-based learning and interactive computer aided language learning in English as a Second Language and foreign languages proved to be most successful in enhancing student performance. This promising multimedia technology is worthy for both school and higher education. An interactive multimedia courseware with storytelling approach supports learning process and the knowledge could be delivered and recognized easier [9]. It is obvious that method of interaction design for enhancing children's cognitive ability is essential [10]. Interactive materials are effective for pre-school students. They can adopt with materials very quickly.

Islam [11] wanted to know the feeling of students when learning materials was playing. All students were very enthusiastic and wanted to see the video more when stopped video. Lukman [12] identifies the commonalities and differences within non-traditional learning methods regarding virtual and real-world environments. A lot of research done where compared online and traditional face-to-face instruction. There are no important difference between face-to-face and online system; any differences that might exist were usually a consequence of teacher involvement and of the commitment of the institution learning process [13]-[20]. Computer aided language learning process is very important and children can take it quickly [21]. There a lot of multimedia institute in Dhaka city and they can produce world quality work. Animation industry in Dhaka city has good prospect although some obstacle need to overcome for success [22]. The requirements for making learning materials are very essential where [23]-[24] described briefly.

The research [25]-[29] has been organized within the five developmental dimensions through which young children learn: social and emotional, language development, physical well-being and motor development, cognitive and general knowledge, and approaches toward learning. And they explore the research on the effectiveness of interactive media to teach disabled, learning disabled and at-risk students. Research review was to provide a description of what future interactive media, especially interactive television, will look like and what their educational potential might be [28]. The multimedia courseware developed for teaching especially pre-school or primary school students. It has a positive impact for promoting learning but designing and developing the materials for their engagement are essential aspects for paying careful attention.

## 3. METHODOLOGY

We have completed our study in two phases.

Firstly, we have developed a solar system learning materials using different multimedia tools like 3D Studio Max. We put animated solar information along with voice of a character inside materials who describes the solar system step by step in Bengali language.

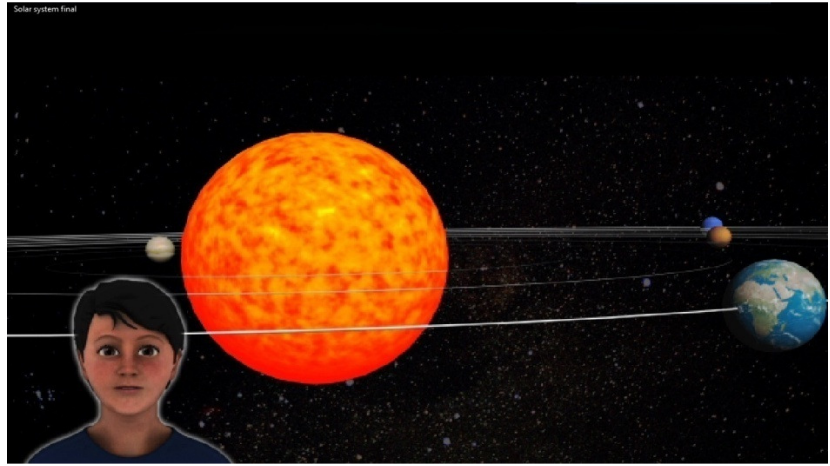

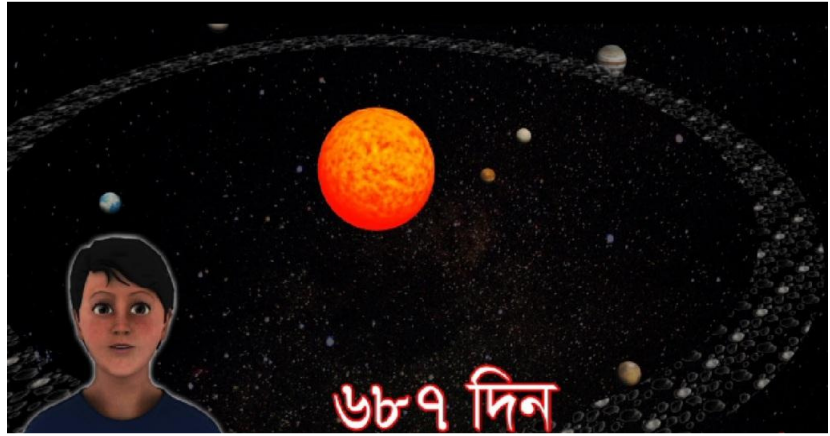

Figure 1 Developed solar system as rendered video using 3ds Max tools

Secondly, we visited one primary school named Ahsania Mohila Mission High School in Dhaka. There were 50 students (aged 4-9 years). We have categorized students in three groups. The number of students in first group was 18, second group was 15 and third group was comprised with 17.

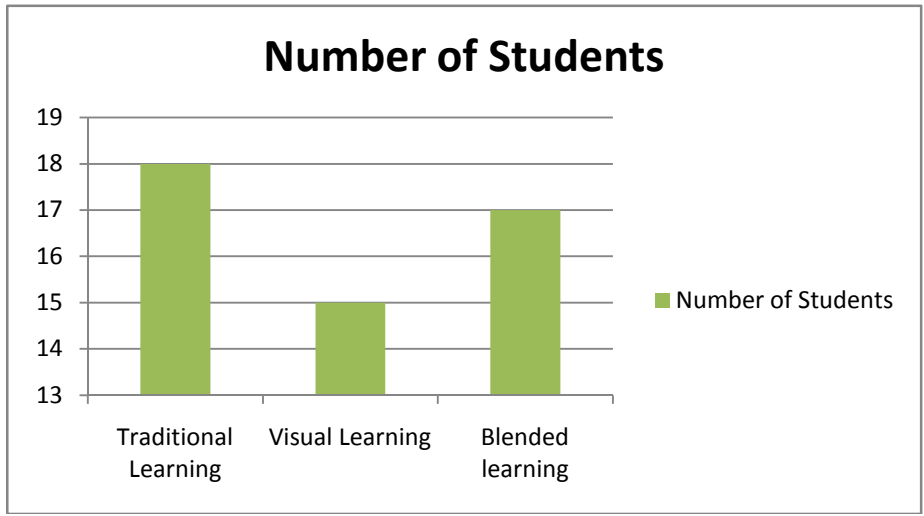

Figure 2 Number of categorized students for testing our system

A teacher taught the first group of students about our solar system using traditional lecture method and then students sat for a test comprised with 15 multiple choice questions. Answer to the questions was related to interactive learning video. Another group of students just watched our interactive learning materials and answered to the 15 questions. The last group of students visualized our materials along with teacher's instructions when needed. Then they sat for answering the same 15 questions. Figures 3 and 4 are shown interactive learning materials for group 2 and 3 students respectively.

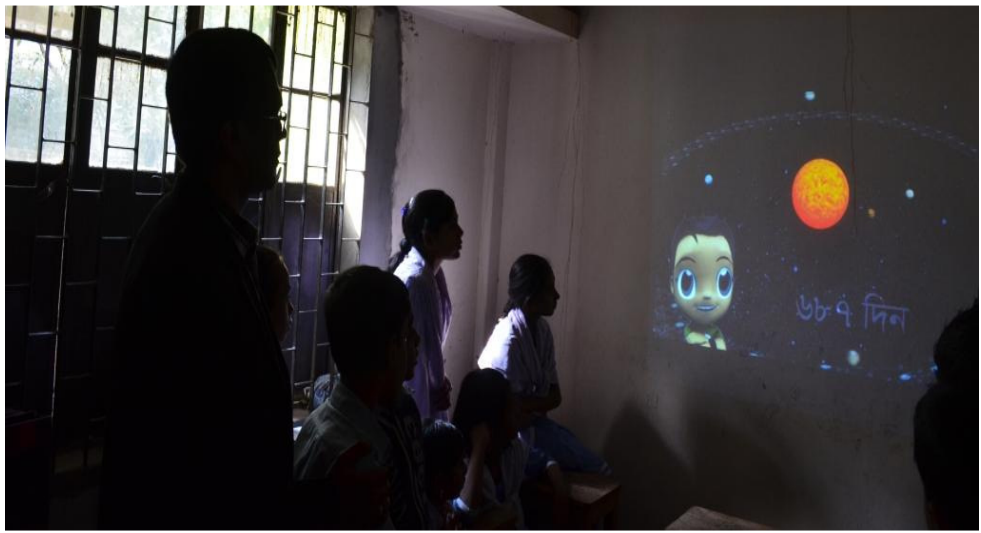

Figure 3 Playing interactive learning materials in front of second group of students in primary School

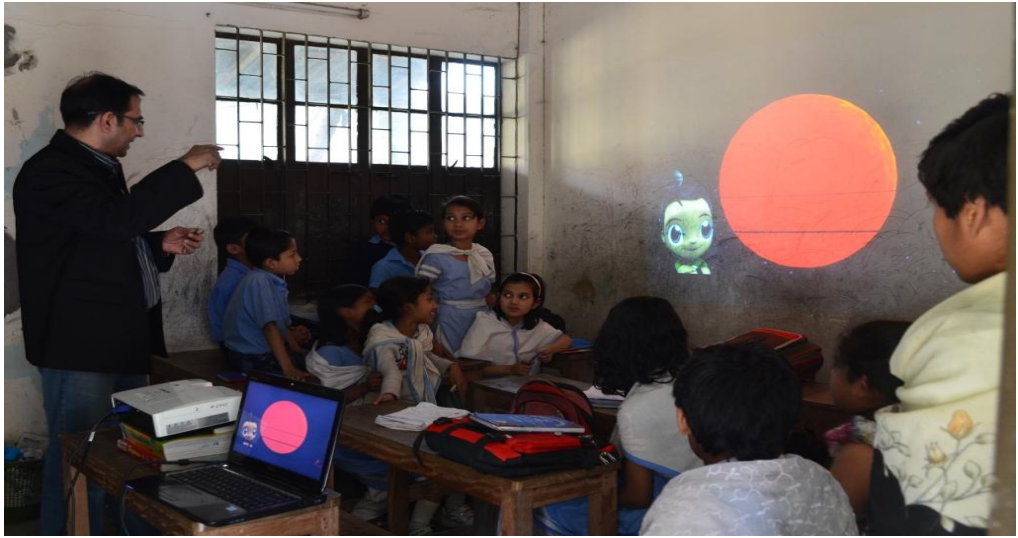

Figure 4 Playing interactive learning materials along with verbal instructions in front of third group of students

## 4. RESULT ANALYSIS

The figure 5 is showing the results of our study. We noticed that eleven students gave 6-10 correct answers by learning traditionally by their teacher. On the other hand, 14 students gave more than 5 correct answers by seeing interactive learning materials once only. But one thing is very noticeable. When students taught using blending system, maximum students gave 11-15 correct answers that are really high compared to other teaching methods.

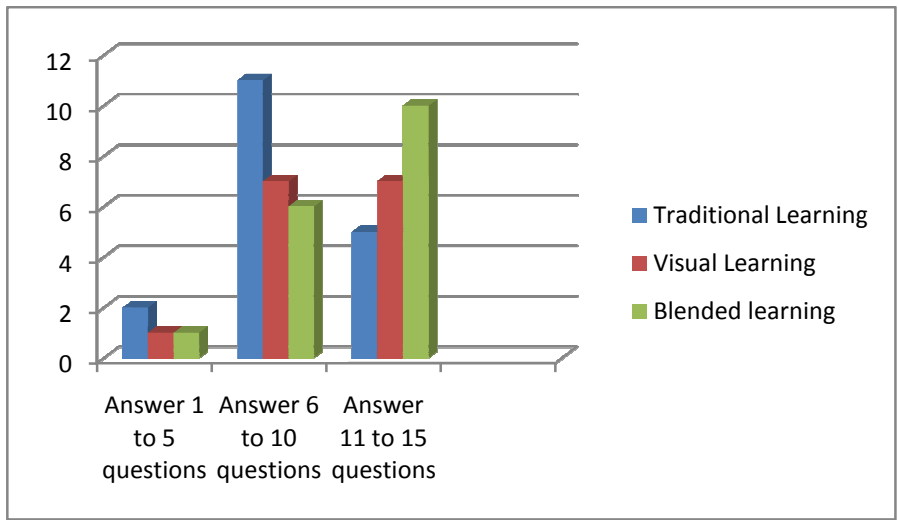

Figure 5 Number of questions answered by students from three groups

We also wanted to know the feeling of students when learning materials were playing. All students were very enthusiastic and wanted to see the video more when we stopped playing our video. The students perceived that the video of the solar system made their ideas clear about the orbit and the planets. Additionally, teacher's commentary was supportive for their understanding. The head of the school, Ms Sirazia Begum said about our developed animated materials for children,

"Definitely these materials will be helpful for students. And they will be quick learner with enjoy and fun."

The expression of the Head teacher indicates that they appreciate this new teaching style for their students.

## 5. CONCLUSIONS

For this study, we developed interactive learning materials on solar system for primary level students. We used different types of multimedia applications and software to develop these materials. We noticed that the impact of interactive learning materials is exclusively high to improve their learning skills and adaptation by blending learning system. Our method is showing the improvement of students learning skills especially when interactive learning materials are used as the main resources by the teacher, although we have no intention the replacement of traditional education system. Just we wanted to introduce a method for promoting learning and quick adaptation with learning materials. This study will be helpful for those who want to work with interactive child education system.

## 6. ACKOWLEDGEMENT

We are thanking to Ms. Sirazia Begum and Ms. Ratna Pervin, Teacher of Ahsania Mohila Mission School for their excellent cooperation for testing and sharing our works with students of the school. We also thank to Mr. Oalid Jahan and Mr. Golam Mawla for their helps in different ways.

**APPENDIX (Questionnaires in Bengali Language):**

# আমাদের সৌরজগৎ

নামঃ _______________________           শ্রেণীঃ _____________           সময়ঃ
১০ মিনিট
সঠিক উত্তরের পাশে টিক চিহ্ন দাও।
১। সৌরজগতের গ্রহ কয়টি?
ক) ৭টি                              খ) ৮টি                              গ) ৯টি
২। সূর্য পৃথিবী থেকে কত গুণ বড়?
ক) ১১ লক্ষ গুন                   খ) ১২ লক্ষ গুন                   গ) ১৩ লক্ষ গুন
৩। সূর্যের সবচেয়ে কাছের গ্রহ কোনটি?
ক) বুধ                              খ) শুক্র                              গ) শনি
৪) আমাদের সৌরজগতের সবচেয়ে ছোট গ্রহ কোনটি?
ক) পৃথিবী                          খ) শুক্র                              গ) বুধ
৫) পৃথিবী সূর্যকে কত দিনে একবার প্রদক্ষিণ করে?
ক) ৩৫৬ দিন                     খ) ৩৬৫ দিন                     গ) ৩৬৮ দিন
৬) মহাশূন্য থেকে পৃথিবীকে কেমন দেখায়?
ক) লাল                            খ) নীল                              গ) সবুজ
৭) সবচেয়ে বড় গ্রহ কোনটি?
ক) পৃথিবী                          খ) মংগল                          গ) বৃহস্পতি
৮) কোন গ্রহকে লাল গ্রহ বলা হয়?
ক) শনি                            খ) মংগল                          গ) শুক্র
৯) সূর্য থেকে সবচেয়ে বেশী দূরে কোন গ্রহটি অবস্থিত?
ক) নেপচুন                        খ) পৃথিবী                          গ) বৃহস্পতি
১০) কোন গ্রহকে শীতলতম গ্রহ বলা হয়?
ক) শনি                            খ) বুধ                              গ) ইউরেনাস
১১) সৌরজগতের চতুর্থ গ্রহ কোনটি?
ক) বৃহস্পতি                       খ) মঙ্গল                           গ) পৃথিবী
১২) মঙ্গল ও বৃহস্পতির মাঝের গ্রহাণুগুলো কত ধরনের হয়ে থাকে?
ক) ২ ধরনের                    খ) ৩ ধরনের                    গ) ৪ ধরনের
১৩) মঙ্গল গ্রহ প্রায় কত দিনে সূর্যকে একবার প্রদক্ষিণ করে?

ক) ৬৩০ দিন   থ) ৭৮৪ দিন   গ) ৬৮৭ দিন

**১৪) কোন গ্রহের রিং সিস্টেম বা বলয় আছে?**
ক) মঙ্গল   থ) বৃহস্পতি   গ) শনি

**১৫) সৌরজগতের আইস জায়ান্ট (ICE GIAMT) হলো-**
ক) ইউরেনাস   থ) শনি   গ) বুধ

## Authors Biography

**Md. Baharul Islam** is a PhD student, Centre of Visual Computing, Multimedia University, Malaysia. His research interests are including image processing, modeling computer animation, computational photography. He received three BEST PAPER AWARD from three international conferences. He has published many articles and presented papers in several conferences. He also received fellowship from Global Institute of science and Technology, Victoria, Australia in 2013.

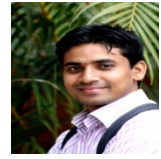

**Arif Ahmed** is Associate professor of Multimedia Technology at Daffodil International University, Bangladesh. He had about 18 years working experience. Currently his research interest is including motion graphics, image processing, stereo image, 3D animation, education system. He started his career as a 3d Animator and Visual Effect developer and served many national and international companies for their 3d visualization works. He created over 200 TV commercials for different satellite channels in Bangladesh. He is the Founder director of AAVA3D, which is the most famous place for research and development of 3d Animation and Visual Arts in Bangladesh. It gives support many local 3D Animation Industry to develop their business.

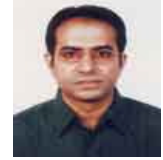

**Dr. Md. Kabirul Islam** is a Professor of Multimedia and Creative Technology Department at Daffodil International University, Bangladesh. He obtained his PhD in E-learning from an Australia. He has more than 25 years working experience with rich publications. His research interests include animation, education, ICT for development, motion analysis etc.

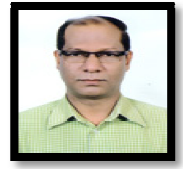

**Abu Kalam Shamsuddin** is a Senior Lecturer of Multimedia and Creative Technology at Daffodil International University, Dhaka, Bangladesh. He received M.F.A from Rajshahi University, Bangladesh in 2008. His research interests are in computer animation, Concept art, Typography. He has work experience in teaching field for three years in Santo-Mariam University of Creative Technology, Dhaka, Bangladesh. He had many published articles and presented papers in different international conferences.

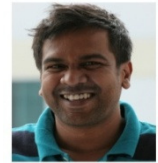